\vsize=9.0in\voffset=1cm
\looseness=2


\message{fonts,}

\font\tenrm=cmr10
\font\ninerm=cmr9
\font\eightrm=cmr8
\font\teni=cmmi10
\font\ninei=cmmi9
\font\eighti=cmmi8
\font\ninesy=cmsy9
\font\tensy=cmsy10
\font\eightsy=cmsy8
\font\tenbf=cmbx10
\font\ninebf=cmbx9
\font\tentt=cmtt10
\font\ninett=cmtt9

\font\ninesl=cmsl9
\font\eightsl=cmsl8

\font\nineit=cmti9
\font\eightit=cmti8

\skewchar\ninei='177 \skewchar\eighti='177
\skewchar\ninesy='60 \skewchar\eightsy='60

\def\eightpoint{\def\rm{\fam0\eightrm} 
\normalbaselineskip=9pt
\normallineskiplimit=-1pt
\normallineskip=0pt

\textfont0=\eightrm \scriptfont0=\sevenrm \scriptscriptfont0=\fiverm
\textfont1=\ninei \scriptfont1=\seveni \scriptscriptfont1=\fivei
\textfont2=\ninesy \scriptfont2=\sevensy \scriptscriptfont2=\fivesy
\textfont3=\tenex \scriptfont3=\tenex \scriptscriptfont3=\tenex
\textfont\itfam=\eightit  \def\it{\fam\itfam\eightit} 
\textfont\slfam=\eightsl \def\sl{\fam\slfam\eightsl} 

\setbox\strutbox=\hbox{\vrule height6pt depth2pt width0pt}%
\normalbaselines \rm}

\def\ninepoint{\def\rm{\fam0\ninerm} 
\textfont0=\ninerm \scriptfont0=\sevenrm \scriptscriptfont0=\fiverm
\textfont1=\ninei \scriptfont1=\seveni \scriptscriptfont1=\fivei
\textfont2=\ninesy \scriptfont2=\sevensy \scriptscriptfont2=\fivesy
\textfont3=\tenex \scriptfont3=\tenex \scriptscriptfont3=\tenex
\textfont\itfam=\nineit  \def\it{\fam\itfam\nineit} 
\textfont\slfam=\ninesl \def\sl{\fam\slfam\ninesl} 
\textfont\bffam=\ninebf \scriptfont\bffam=\sevenbf
\scriptscriptfont\bffam=\fivebf \def\bf{\fam\bffam\ninebf} 
\textfont\ttfam=\ninett \def\tt{\fam\ttfam\ninett} 

\normalbaselineskip=11pt
\setbox\strutbox=\hbox{\vrule height8pt depth3pt width0pt}%
\let \smc=\sevenrm \let\big=\ninebig \normalbaselines
\parindent=1em
\rm}

\def\tenpoint{\def\rm{\fam0\tenrm} 
\textfont0=\tenrm \scriptfont0=\ninerm \scriptscriptfont0=\fiverm
\textfont1=\teni \scriptfont1=\seveni \scriptscriptfont1=\fivei
\textfont2=\tensy \scriptfont2=\sevensy \scriptscriptfont2=\fivesy
\textfont3=\tenex \scriptfont3=\tenex \scriptscriptfont3=\tenex
\textfont\itfam=\nineit  \def\it{\fam\itfam\nineit} 
\textfont\slfam=\ninesl \def\sl{\fam\slfam\ninesl} 
\textfont\bffam=\ninebf \scriptfont\bffam=\sevenbf
\scriptscriptfont\bffam=\fivebf \def\bf{\fam\bffam\tenbf} 
\textfont\ttfam=\tentt \def\tt{\fam\ttfam\tentt} 

\normalbaselineskip=11pt
\setbox\strutbox=\hbox{\vrule height8pt depth3pt width0pt}%
\let \smc=\sevenrm \let\big=\ninebig \normalbaselines
\parindent=1em
\rm}

\message{fin format jgr}

\vskip 4 mm
\magnification=1200
\font\Bbb=msbm10
\def\R{\hbox{\Bbb R}}
\def\C{\hbox{\Bbb C}}
\def\S{\hbox{\Bbb S}}
\def\pa{\partial}
\def\ep{\varepsilon}
\def\b{\backslash}
\def\v{\varphi}
\vskip 4 mm
\centerline{\bf Formulas and equations for finding scattering data}
\centerline{\bf from the Dirichlet-to-Neumann map with nonzero
background potential}
\vskip 4 mm

\centerline{\bf R. G. Novikov}
\vskip 4 mm
\noindent
{\ninerm CNRS, Laboratoire de Math\'ematiques Jean Leray (UMR 6629),
Universit\'e de Nantes, BP 92208,}

\noindent
{\ninerm F-44322, Nantes cedex 03, France}

\noindent
{\ninerm e-mail: novikov@math.univ-nantes.fr}

\vskip 4 mm
{\bf Abstract}

For the Schr\"odinger equation at fixed energy with a potential
supported in a bounded domain we give formulas and equations for finding
scattering data from the Dirichlet-to-Neumann map with nonzero
background potential. For the case of zero background potential these
results were obtained in [R.G.Novikov, Multidimensional inverse spectral
problem for the equation $-\Delta\psi + (v(x)-Eu(x))\psi=0$, Funkt. Anal. i
Ego Prilozhen 22(4), pp.11-22, (1988)].

\vskip 4 mm
{\bf 1.Introduction}

Consider the Schr\"odinger equation
$$-\Delta\psi+v(x)\psi=E\psi,\ \ x\in D,\eqno(1.1)$$
where
$$\eqalign{
&D\ \ {\rm is\ an\ open\ bounded\ domain\ in}\ \ \R^d,\ \ d\ge 2,\cr
&{\rm with}\ \ \pa D\in C^2,\cr}\eqno(1.2a)$$
$$v\in L^{\infty}(D).\eqno(1.2b)$$
We also assume that
$$\eqalign{
&E\ \ {\rm is\ not\ a\ Dirichlet\ eigenvalue\ for}\cr
&{\rm the\ operator}\ \ -\Delta+v\ \ {\rm in}\ \ D.\cr}\eqno(1.3)$$
Consider the map $\Phi(E)$ such that
$${\pa\psi\over \pa\nu}\big|_{\pa D}=\Phi(E)\bigl(\psi\big|_{\pa D}\bigr)
\eqno(1.4)$$
for all sufficiently regular solutions of (1.1) in $\bar D=D\cup\pa D$, for
example, for all $\psi\in H^1(D)$ satisfying (1.1),
where $\nu$ is the outward normal to $\pa D$.
 The map $\Phi(E)$ is
called the Dirichlet-to-Neumann map for equation (1.1).

Consider the Schr\"odinger equation
$$-\Delta\psi+v(x)\psi=E\psi,\ \ x\in\R^d,\eqno(1.5)$$
where
$$\rho^{d+\ep}v\in L^{\infty}(\R^d),\ \ d\ge 2,\ \ {\rm for\ some}\ \ \ep>0,
\eqno(1.6)$$
where $\rho$ denotes the multiplication operator by the function $\rho(x)=
1+|x|$. For equation (1.5) we consider the functions $\psi^+$ and $f$ of the
classical scattering theory and the Faddeev functions $\psi$, $h$,
$\psi_{\gamma}$, $h_{\gamma}$ (see, for example, [F1], [F2], [F3], [HN],
[Ne]).

The functions $\psi^+$ and $f$ are defined as follows:
$$\eqalignno{
&\psi^+(x,k)=e^{ikx}+\int_{\R^d}G^+(x-y,k)v(y)\psi^+(y,k)dy, &(1.7)\cr
&G^+(x,k)=-\bigl({1\over 2\pi}\bigr)^d\int_{\R^d}
{e^{i\xi x}d\xi\over {\xi^2-k^2-i0}},&(1.8)\cr}$$
where $x,k\in\R^d$, $k^2>0$ (and at fixed $k$ the formula (1.7) is an
equation for $\psi^+$ in $L^{\infty}(\R^d)$);
$$f(k,l)=\bigl({1\over 2\pi}\bigr)^d\int_{\R^d}e^{-ilx}v(x)\psi^+(x,k)dx,
\eqno(1.9)$$
where $k,l\in\R^d$, $k^2>0$. Here $\psi^+(x,k)$ satisfies (1.5) for
$E=k^2$ and describes scattering of the plane waves $e^{ikx}$; $f(k,l)$,
$k^2=l^2$, is the scattering amplitude for equation (1.5) for $E=k^2$.
The equation (1.7) is called the Lippman-Schwinger integral equation.

The functions $\psi$ and $h$ are defined as follows:
$$\eqalignno{
&\psi(x,k)=e^{ikx}+\int_{\R^d}G(x-y,k)v(y)\psi(y,k)dy, &(1.10)\cr
&G(x,k)=e^{ikx}g(x,k),\ \
g(x,k)=-\bigl({1\over 2\pi}\bigr)^d\int_{\R^d}
{e^{i\xi x}d\xi\over {\xi^2+2k\xi}},&(1.11)\cr}$$
where $x\in\R^d$, $k\in\C^d\b\R^d$ (and at fixed $k$ the formula (1.10) is
an equation for $\psi=e^{ikx}\mu(x,k)$, where $\mu$ is sought in
$L^{\infty}(\R^d))$;
$$h(k,l)=\bigl({1\over 2\pi}\bigr)^d\int_{\R^d}e^{-ilx}v(x)\psi(x,k)dx,
\eqno(1.12)$$
where $k,l\in\C^d\b\R^d$, $Im\,k=Im\,l$. Here $\psi(x,k)$ satisfies
(1.5) for $E=k^2$, and $\psi$, $G$ and $h$ are (nonanalytic)
continuations of $\psi^+$, $G^+$ and $f$ to the complex domain. In particular,
 $h(k,l)$ for $k^2=l^2$ can be considered as the "scattering" amplitude
in the complex domain for equation (1.5) for $E=k^2$.
The functions $\psi_{\gamma}$ and $h_{\gamma}$ are defined as follows:
$$\psi_{\gamma}(x,k)=\psi(x,k+i0\gamma),\ \ h_{\gamma}(k,l)=h(k+i0\gamma,
l+i0\gamma),\eqno(1.13)$$
where $x,k,l,\gamma\in\R^d$, $|\gamma|=1$. Note that
$$\psi^+(x,k)=\psi_{k/|k|}(x,k),\ \ f(k,l)=h_{k/|k|}(k,l),\eqno(1.14)$$
where $x,k,l\in\R^d$, $|k|>0$.

We consider $f(k,l)$ and $h_{\gamma}(k,l)$, where $k,l,\gamma\in\R^d$,
$k^2=l^2=E$, $\gamma^2=1$, and $h(k,l)$, where $k,l\in\C^d\b\R^d$,
$Im\,k=Im\,l$, $k^2=l^2=E$, as scattering data $S_E$ for equation (1.5) at
fixed $E\in ] 0, +\infty [$. We consider $h(k,l)$, where $k,l\in\C^d\b\R^d$,
$Im\,k=Im\,l$, $k^2=l^2=E$, as scattering data $S_E$ for equation (1.5) at
fixed $E\in\C\b ] 0, +\infty [$.

Let $D$ be a fixed domain satisfying (1.2a). Let
$$v\in L^{\infty}(D)\ \ {\rm and}\ \ v\equiv 0\ \ {\rm on}\ \ \R^d\b\bar D.
\eqno(1.15)$$
For $v$ of (1.15) we consider the Dirichlet-to-Neumann map $\Phi(E)$ for
equation (1.1) and the scattering data $S_E$ for equation (1.5).

In the present work we continue studies of [No1] on the following
inverse boundary value problem for equation (1.1):

{\bf Problem 1.}
Find $v$ (in (1.1)) from $\Phi(E)$ (where $E$ is fixed or belongs to some
set).

More precisely, we develop formulas and equations of [No1] which reduce
Problem 1 to the following inverse scattering problem for equation (1.5):

{\bf Problem 2.}
Find $v$ (in (1.5)) from $S(E)$ (where $E$ is fixed or belongs to some set).

Concerning results given in the literature on Problem 1, see [SU], [No1],
[A], [NSU], [Na1], [Na2], [M] and references therein. Concerning results
given in the literature on Problem 2, see [BC], [HN], [No2], [No3], [IS],
[GN], [No4], [No5], [E], [Ch], [BBMRS], [BMR] and references therein.

The main results of the present work consist of Theorem 1 and Propositions 1
and 2
of Section 2. In these results we consider for fixed $E$ two potentials
$v_0$ and $v$ satisfying (1.15) and (1.3). In Theorem 1 and Proposition 1
we give formulas and
equations for finding $S_E$ from $\Phi(E)-\Phi_0(E)$ and from (some functions
found from) $v_0$, where $S_E$ and $\Phi(E)$ correspond to $v$ and $\Phi_0(E)$
 corresponds to $v_0$. In Proposition 2 we give a result on the solvability
of equations of Theorem 1.

For the case when $v_0\equiv 0$, Theorem 1 and Propositions 1 and 2 were
obtained
for the first time in [No1] (see also [HN], [Na1], [Na2]), where using
these results Problem 1 was reduced to Problem 2. For the case when the best
known approximation $v_0$ to $v$ of Problem 1 is not identically zero
(and especially when $v_0$ is really close to $v$), results of the present
work reduce Problem 1 to Problem 2 in a more stable way than it was done
in [No1].

Note that generalizing results of [No1] to the case of nonzero background
potential $v_0$ we used essentially results of [No4]. The main results of
the present work  are presented in detail in Section 2.

Note, finally, that by the present work we start a development of new exact
reconstruction algorithms for Problem 1.

\vskip 4 mm
{\bf 2. Main results}

To formuate our results we need to introduce some additional notations.
Consider (under the assumption (1.6)) the sets ${\cal E}$, ${\cal E}_{\gamma}$,
${\cal E}^+$ defined as follows:
$$\eqalign{
&{\cal E}=\{\zeta\in\C^d\b\R^d\ :\ \ {\rm equation}\ \ (1.10)\ \ {\rm for}\ \
k=\zeta\ \ {\rm is\ not}\cr
&{\rm uniquely\ solvable\ for}\ \ \psi=e^{ikx}\mu\ \ {\rm with}\ \
\mu\in L^{\infty}(\R^d)\},\cr}\eqno(2.1a)$$
$$\eqalign{
&{\cal E}_{\gamma}=\{\zeta\in\R^d\b 0\ :\ \ {\rm equation}\ \ (1.10)\ \ {\rm for}\ \
k=\zeta+i0\gamma\ \ {\rm is\ not}\cr
&{\rm uniquely\ solvable\ for}\ \
\psi\in L^{\infty}(\R^d)\},\ \ \gamma\in\S^{d-1},\cr}\eqno(2.1b)$$
$$\eqalign{
&{\cal E}^+=\{\zeta\in\R^d\b 0\ :\ \ {\rm equation}\ \ (1.7)\ \ {\rm for}\
k=\zeta\ \ {\rm is\ not}\cr
&{\rm uniquely\ solvable\ for}\ \
\psi^+\in L^{\infty}(\R^d)\}.\cr}\eqno(2.1c)$$

Note that ${\cal E}^+$ is a well-known set of the classical scattering theory
for equation (1.5) and that ${\cal E}^+=\emptyset$ for real-valued $v$ satisfying
(1.6)
(see, for example, [Ne]). The sets ${\cal E}$ and ${\cal E}_{\gamma}$ were
considered
for the first time in [F1], [F2], [F3]. Concerning the properties of
${\cal E}$
and ${\cal E}_{\gamma}$, see [F3], [HN], [LN], [Ne], [We], [Na2], [No4],
[No6].

Consider (under the assumptions (1.6)) the functions $R$, $R_{\gamma}$,
$R^+$ defined as follows:
$$R(x,y,k)=G(x-y,k)+\int_{\R^d}G(x-z,k)v(z)R(z,y,k)dz,\eqno(2.2)$$
where $x,y\in\R^d$, $k\in\C^d\b\R^d$, $G$ is defined by (1.11), and at
fixed $y$ and $k$ the formula (2.2) is an equation for
$$R(x,y,k)=e^{ik(x-y)}r(x,y,k),\eqno(2.3)$$
where $r$ is sought with the properties
$$\eqalignno{
&r(\cdot,y,k)\ \ {\rm is\ continuous\ on}\ \ \R^d\b y,&(2.4a)\cr
&r(x,y,k)\to 0\ \ {\rm as}\ \ |x|\to\infty,&(2.4b)\cr}$$
$$\eqalign{
&r(x,y,k)=O(|x-y|^{2-d})\ \ {\rm as}\ \ x\to y\ \ {\rm for}\ \ d\ge 3,\cr
&r(x,y,k)=O(|\ln\,|x-y||)\ \ {\rm as}\ \ x\to y\ \ {\rm for}\ \ d=2;\cr}
\eqno(2.4c)$$
$$R_{\gamma}(x,y,k)=R(x,y,k+i0\gamma),\eqno(2.5)$$
where $x,y\in\R^d$, $k\in\R^d\b 0$, $\gamma\in\S^{d-1}$;
$$R^+(x,y,k)=R_{k/|k|}(x,y,k),\eqno(2.6)$$
where $x,y\in\R^d$, $k\in\R^d\b 0$. Note that $R(x,y,k)$, $R_{\gamma}(x,y,k)$
and $R^+(x,y,k)$ (for their domains of definition in $k$ and $\gamma$)
satisfy the equation
$$(\Delta+E-v(x))R(x,y,k)=\delta(x-y),\ \ x\in\R^d,\ \ y\in\R^d,\ \ E=k^2.
\eqno(2.7)$$
The function $R^+(x,y,k)$ (defined by means of (2.2) for $k\in\R^d\b 0$
with $G$ replaced by $G^+$ of (1.8)) is well-known in the scattering theory
for equations (1.5), (2.7). In particular, this function  describes
scattering of the spherical waves $G^+(x-y,k)$ generated by a source at $y$.
Apparently, the functions $R$ and $R_{\gamma}$ were considered for the
first time in [No5].

Note that under the assumption (1.6): equation (2.2) at fixed $y$ and $k$
is uniquely solvable for $R$ with the properties (2.3), (2.4) if and only if
$k\in\C^d\b (\R^d\cup {\cal E})$; equation (2.2) with $k=\zeta+i0\gamma$,
$\zeta\in\R^d\b 0$, $\gamma\in\S^{d-1}$, at fixed $y$, $\zeta$ and $\gamma$
is uniquely solvable for $R_{\gamma}$ if and only if $\zeta\in\R^d\b
(0\cup {\cal E}_{\gamma})$; equation (2.2) with $k=\zeta+i0{\zeta/|\zeta|}$,
$\zeta\in\R^d\b 0$, at fixed $y$ and $\zeta$ is uniquely solvable for
$R^+$ if and only if $\zeta\in\R^d\b (0\cup {\cal E}^+)$.

For $v$ of (1.15) we consider the map $\Phi(E)$ defined by means of (1.4),
the functions $\psi^+$, $f$, $\psi$, $h$, $\psi_{\gamma}$, $h_{\gamma}$ and
$R^+$, $R$, $R_{\gamma}$ defined by means of (1.7)-(1.13) and (2.2)-(2.6)
and the sets ${\cal E}$, ${\cal E}_{\gamma}$, ${\cal E}^+$ defined by (2.1).
The Schwartz
kernel of the integral operator $\Phi(E)$ will be denoted by
$\Phi(x,y,E)$, where $x,y\in \pa D$.

\vskip 2 mm
{\bf Theorem 1.}
{\it Let} $D$ {\it satisfying} (1.2a) {\it and} $E$ {\it be fixed. Let}
$v^0$ {\it and} $v$ {\it be two potentials satisfying} (1.15), (1.3). {\it
Let} $\Phi$, $\psi^+$, $f$, $\psi$, $h$, $\psi_{\gamma}$, $h_{\gamma}$,
$R^+$, $R$, $R_{\gamma}$, ${\cal E}$, ${\cal E}_{\gamma}$, ${\cal E}^+$
{\it correspond to}
$v$ {\it (as defined above) and} $\Phi^0$, $\psi^{+,0}$, $f^0$, $\psi^0$,
$h^0$, $\psi^0_{\gamma}$, $h^0_{\gamma}$, $R^{+,0}$, $R^0$, $R^0_{\gamma}$,
${\cal E}^0$, ${\cal E}^0_{\gamma}$, ${\cal E}^{+,0}$ {\it correspond to}
$v^0$ {\it (as defined above with} $v=v^0$).
{\it Then the following formulas hold:}
$$h(k,l)=h^0(k,l)+\bigl({1\over 2\pi}\bigr)^d\int_{\pa D}\int_{\pa D}
\psi^0(x,-l)(\Phi-\Phi^0)(x,y,E)\psi(y,k)dydx \eqno(2.8)$$
{\it for} $k,l\in\C^d\b (\R^d\cup {\cal E}^0\cup {\cal E})$, $Im\,k=Im\,l$,
$k^2=l^2=E$,
$$\eqalignno{
&\psi(x,k)=\psi^0(x,k)+\int_{\pa D}A(x,y,k)\psi(y,k)dy,\ \ x\in\pa D,&(2.9a)
\cr
&A(x,y,k)=\int_{\pa D}R^0(x,z,k)(\Phi-\Phi^0)(z,y,E)dz,\ \ x,y\in\pa D,
&(2.9b)\cr}$$
{\it for} $k\in\C^d\b (\R^d\cup {\cal E}^0\cup {\cal E})$, $k^2=E$;
$$h_{\gamma}(k,l)=h^0_{\gamma}(k,l)+\bigl({1\over 2\pi}\bigr)^d\int_{\pa D}
\int_{\pa D}\psi^0_{-\gamma}(x,-l)(\Phi-\Phi^0)(x,y,E)
\psi_{\gamma}(y,k)dydx \eqno(2.10)$$
{\it for} $k,l\in\R^d\b (0\cup {\cal E}^0_{\gamma}\cup {\cal E}_{\gamma})$,
$\gamma\in\S^{d-1}$, $k^2=l^2=E$, $k\gamma=l\gamma$,
$$\eqalignno{
&\psi_{\gamma}(x,k)=\psi^0_{\gamma}(x,k)+\int_{\pa D}A_{\gamma}(x,y,k)
\psi_{\gamma}(y,k)dy,\ \ x\in\pa D,&(2.11a)\cr
&A_{\gamma}(x,y,k)=\int_{\pa D}R^0_{\gamma}(x,z,k)(\Phi-\Phi^0)(z,y,E)dz,\ \
x,y\in\pa D,&(2.11b)\cr}$$
{\it for} $k\in\R^d\b (0\cup {\cal E}^0_{\gamma}\cup {\cal E}_{\gamma})$,
$\gamma\in\S^{d-1}$, $k^2=E$;
$$f(k,l)=f^0(k,l)+\bigl({1\over 2\pi}\bigr)^d\int_{\pa D}\int_{\pa D}
\psi^{+,0}(x,-l)(\Phi-\Phi^0)(x,y,E)\psi^+(y,k)dydx \eqno(2.12)$$
{\it for} $k,l\in\R^d\b (0\cup {\cal E}^{+,0}\cup {\cal E}^+)$, $k^2=l^2=E$,
$$\eqalignno{
&\psi^+(x,k)=\psi^{+,0}(x,k)+\int_{\pa D}A^+(x,y,k)\psi^+(y,k)dy,\ \
x\in\pa D,&(2.13a)\cr
&A^+(x,y,k)=\int_{\pa D}R^+(x,z,k)(\Phi-\Phi^0)(z,y,E)dz,\ \ x,y\in\pa D,
&(2.13b)\cr}$$
{\it for} $k\in\R^d\b (0\cup {\cal E}^{+,0}\cup {\cal E}^+)$, $k^2=E$.

Note that in Theorem 1 $dx$ and $dy$ denote the standard measure on
$\pa D$ in $\R^d$.

Note that in the formula (2.10) for $h_{\gamma}(k,l)$ there is the
additional restriction: $k\gamma=l\gamma$. To extend (2.10) to the general
case, consider $\psi_{\gamma}(x,k,l)$ defined as follows:
$$\eqalignno{
&\psi_{\gamma}(x,k,l)=e^{ilx}+\int_{\R^d}G_{\gamma}(x-y,k)v(y)
\psi_{\gamma}(y,k,l)dy,&(2.14a)\cr
&G_{\gamma}(x,k)=G(x,k+i0\gamma),&(2.14b)\cr}$$
where $x,k,l\in\R^d$, $k^2=l^2>0$, $\gamma\in\S^{d-1}$ and (2.14) at fixed
$\gamma$, $k$, $l$ is an equation for $\psi_{\gamma}(\cdot,k,l)$ in
$L^{\infty}(\R^d)$.

\vskip 2 mm
{\bf Proposition 1.}
{\it Let the assumptions of Theorem 1 be valid. In addition, let}
$\psi_{\gamma}(x,k,l)$ {\it correspond to} $v$ {\it and}
$\psi^0_{\gamma}(x,k,l)$ {\it correspond to} $v^0$. {\it Then}
$$h_{\gamma}(k,l)=h^0_{\gamma}(k,l)+\bigl({1\over 2\pi}\bigr)^d
\int_{\pa D}\int_{\pa D}\psi^0_{-\gamma}(x,-k,-l)(\Phi-\Phi^0)(x,y,E)
\psi_{\gamma}(y,k)dydx \eqno(2.15)$$
{\it for} $\gamma\in\S^{d-1}$, $k\in\R^d\b (0\cup {\cal E}^0_{\gamma}
\cup {\cal E}_{\gamma})$, $l\in\R^d$, $k^2=l^2=E$.

Note that (see [F3], [No4])
$$G_{\gamma}(x,k)=G_{\gamma}(x,l)\ \ {\rm for}\ \ x,k,l\in\R^d,\ \
\gamma\in\S^{d-1},\ \ k^2=l^2>0,\ \ k\gamma=l\gamma.\eqno(2.16)$$
Therefore,
$$\psi_{\gamma}(x,k,l)=\psi_{\gamma}(x,l,l)=\psi_{\gamma}(x,l)\ \ {\rm for}\ \
 x,k,l\in\R^d,\ \ \gamma\in\S^{d-1},\ \ k^2=l^2>0,\ \ k\gamma=l\gamma.
\eqno(2.17)$$
Therefore, (2.15), under the additional restriction $k\gamma=l\gamma$
is reduced  to (2.10).

Suppose that $v$ is unknown, but $v^0$ and $\Phi(E)-\Phi^0(E)$ are known
($v^0$ is considered as the best known approximation to $v$). Then Theorem 1
and Proposition 1 (and equations and formulas (1.7)-(1.13), (2.2)-(2.6),
(2.14) for finding $\psi^{+,0}$, $f^0$, $\psi^0$, $h^0$, $\psi^0_{\gamma}$,
$h^0_{\gamma}$, $R^{+,0}$, $R^0$, $R^0_{\gamma}$ from $v^0$) give a method
for finding the scattering data $S_E$ (defined in the introduction) for
$v$ from the background potential $v^0$ and the difference
$\Phi(E)-\Phi^0(E)$. In addition, (2.9a), (2.11a), (2.13a) at fixed $k$ and
$\gamma$ are linear integral equations for finding $\psi$, $\psi_{\gamma}$,
$\psi^+$ on $\pa D$ from $\psi^0$, $\psi^0_{\gamma}$, $\psi^{+,0}$ on
$\pa D$ and $A$, $A_{\gamma}$, $A^+$ on $\pa D\times\pa D$ (where $A$,
$A_{\gamma}$, $A^+$ are given by (2.9b), (2.11b), (2.13b)).

\vskip 2 mm
{\bf Proposition 2.}
{\it Under the assumptions of Theorem 1, equation} (2.9a) {\it at fixed}
$k\in\C^d\b (\R^d\cup {\cal E}^0)$, $k^2=E$, {\it equation} (2.11a) {\it at
fixed} $\gamma\in\S^{d-1}$ {\it and} $k\in\R^d\b (0\cup {\cal E}^0_{\gamma})$,
 $k^2=E$, {\it and equation} (2.13a) {\it at fixed} $k\in\R^d\b (0\cup
{\cal E}^{+,0})$ {\it are Fredholm linear integral equations of the second
kind for} $\psi$, $\psi_{\gamma}$ {\it and} $\psi^+$ {\it (respectively) in}
$L^{\infty}(\pa D)$ {\it and are uniquely solvable (in this space) if and
only if} $k\notin {\cal E}$ {\it for} (2.9a), $k\notin {\cal E}_{\gamma}$
{\it for} (2.11a) {\it and} $k\notin {\cal E}^+$ {\it for} (2.13a).

Note that
$$\psi(x,k)=e^{ikx}\mu(x,k),\ \ \psi^0(x,k)=e^{ikx}\mu^0(x,k),\ \ x\in\R^d,\
\ k\in\C^d\b\R^d,\eqno(2.18)$$
where $e^{ikx}$ is an exponentially increasing factor and
$$\mu(x,k)\to 1,\ \ \mu^0(x,k)\to 1\ \ {\rm as}\ \ |k|\to\infty \eqno(2.19)$$
for $k^2=E$ at fixed $E$, where $|k|=\sqrt{(Re\,k)^2+(Im\,k)^2}$.
Therefore, it is convenient to write equation (2.9) of Theorem 1 as follows:
$$\eqalignno{
&\mu(x,k)=\mu^0(x,k)+\int_{\pa D}B(x,y,k)\mu(y,k)dy,\ \ x\in\pa D,&(2.20a)\cr
&B(x,y,k)=\int_{\pa D}r^0(x,z,k)e^{-ikz}(\Phi-\Phi^0)(z,y,E)dz\,e^{iky},\ \
x,y\in\pa D,&(2.20b)\cr}$$
for $k\in\C^d\b (\R^d\cup {\cal E}^0\cup {\cal E})$, where $r^0$ and
$R^0$ are related by (2.3).

Theorem 1 and Proposition 1 reduce Problem 1 to Problem 2 (these problems are
formulated in the introduction).

For the case when $v^0\equiv 0$, Theorem 1 and Propositions 1 and 2 were
obtained in [No1] (see also [Na1], [Na2]). Note that the basic results of
[No1], in particular formula (2.8) and equation (2.9) for $v^0\equiv 0$ and
$d=3$, were presented already in the survey given in [HN].

For the case when the best known approximation $v^0$ to $v$ is not
identically zero (and especially when $v^0$ is really close to $v$) Theorem 1
 and Proposition 1 give a more convenient (in particular, for the stability
analysis) method for reducing Problem 1 to Problem 2 than in [No1]. To
explain this more precisely, consider, in particular, the integral operators
$B(k)$, $A_{\gamma}(k)$, $A^+(k)$ (with the Schwartz kernels $B(x,y,k)$,
$A_{\gamma}(x,y,k)$, $A^+(x,y,k)$) of equations (2.20a), (2.11a), (2.13a).
To have a simple and stable (with respect to small errors in $\Phi(E)$)
method for solving equations (2.20a) (for fixed
$k\in\C^d\b (\R^d\cup {\cal E}^0)$, $k^2=E$), (2.11a) (for fixed
$\gamma\in\S^{d-1}$ and $k\in\R^d\b (0\cup {\cal E}^0_{\gamma})$, $k^2=E$)
and (2.13a) (for fixed $k\in\R^d\b (0\cup {\cal E}^{+,0})$, $k^2=E$) it is
important to have that
$$\|B(k)\|<\eta,\ \ \|A_{\gamma}(k)\|<\eta,\ \ \|A^+(k)\|<\eta,\eqno(2.21)$$
respectively, for some $\eta<1$, where $\|A\|$ is the norm of an operator
$A$ (for example) in $L^{\infty}(\pa D)$. In this case equations (2.20a),
(2.11a), (2.13a) are uniquely solvable by the method of successive
approximations. In addition, if $\eta\ll 1$, then (2.20a), (2.11a), (2.13a)
can be solved in the first approximation as
$$\mu(x,k)\approx\mu^0(x,k),\ \ \psi_{\gamma}(x,k)\approx\psi^0_{\gamma}(x,k),
\ \ \psi^+(x,k)\approx\psi^{+,0}(x,k) \eqno(2.22)$$
and $h$, $h_{\gamma}$, $f$ can be determined in the first (nontrivial)
approximation as
$$h(k,l)\approx h^0(k,l)+\bigl({1\over 2\pi}\bigr)^d
\int_{\pa D}\int_{\pa D}\psi^0(x,-l)(\Phi-\Phi^0)(x,y,E)
\psi^0(y,k)dydx \eqno(2.23a)$$
for $l\in\C^d\b\R^d$, $Im\,l=Im\,k$, $l^2=k^2=E$ (and for $k$ of $B(k)$ of
(2.21)),
$$h_{\gamma}(k,l)\approx h^0_{\gamma}(k,l)+\bigl({1\over 2\pi}\bigr)^d
\int_{\pa D}\int_{\pa D}\psi^0_{-\gamma}(x,-k,-l)(\Phi-\Phi^0)(x,y,E)
\psi^0_{\gamma}(y,k)dydx \eqno(2.23b)$$
for $l\in\R^d\b {\cal E}^0_{\gamma}$, $l^2=k^2=E$ (and for $k$, $\gamma$ of
$A_{\gamma}(k)$ of (2.21)),
$$f(k,l)\approx f^0(k,l)+\bigl({1\over 2\pi}\bigr)^d\int_{\pa D}\int_{\pa D}
\psi^{+,0}(x,-l)(\Phi-\Phi^0)(x,y,E)\psi^{+,0}(y,k)dydx \eqno(2.23c)$$
for $l\in\R^d$, $l^2=k^2=E$ (and for $k$ of $A^+(k)$ of (2.21)).

Note that the  direct problem of finding $\psi^{+,0}$, $f^0$, $\psi^0$,
$h^0$, $\psi^0_{\gamma}$, $h^0_{\gamma}$, $R^{+,0}$, $R^0$, $R^0_{\gamma}$
(involved into (2.8)-(2.13), (2.15), (2.20)) from $v^0$ is (relatively)
well understood (in comparison with the problem of solving (2.20a),
(2.11a), (2.13a) without the assumptions (2.21)) and is sufficiently stable.

Note that in (2.20a), apparently, unfortunately, almost always
$$\|B(k)\|\to\infty\ \ {\rm (exponentially\ fast)\ as}\ \ |k|\to\infty
\eqno(2.24)$$
for $k\in\C^d$, $k^2=E$ at fixed $E$, in spite of (2.19).

To have (2.21) for $B(k)$ for maximally large domain in
$k\in\C^d\b\R^d$, $k^2=E$, and when $E>0$ for $A_{\gamma}(k)$ for
maximally large domain in $\gamma\in\S^{d-1}$ and $k\in\R^d$, $k^2=E$, and
for $A^+(k)$ for $k\in\R^d$, $k^2=E$, it is important to have that
$\|\Phi(E)-\Phi^0(E)\|$ is as small as possible. The smallness of
$\|\Phi(E)-\Phi^0(E)\|$ follows from the closeness of $v^0$ to $v$ (for
example) in $L^{\infty}(D)$ for fixed $D$, $v$ and $E$, under the
conditions (1.2), (1.3).

As soon as Problem 1 is reduced to Problem 2, one can use for solving
Problem1 methods of [HN], [No2], [Na2], [IS], [No4], [No5], [E], [BBMRS],
[BMR] (and further references given therein).

\vskip 2 mm
{\bf 3. Proofs of Theorem 1 and Propositions 1 and 2}
\vskip 2 mm
For the case when $v^0\equiv 0$, Theorem 1 and Propositions 1 and 2 were
proved in [No1]. In this section we generalize these proofs of [No1] to the
case of nonzero background potential $v^0$. To this end we use, in
particular, some results of [A] and [No4].

\vskip 2 mm
{\bf Proof of Theorem 1.}
We proceed from the following formulas and equations (being valid under the
assumption (1.6) on $v^0$ and $v$):
$$h(k,l)=h^0(k,l)+\bigl({1\over 2\pi}\bigr)^d\int_{\R^d}\psi^0(x,-l)
(v(x)-v^0(x))\psi(x,k)dx \eqno(3.1)$$
for $k,l\in\C^d\b (\R^d\cup {\cal E}^0\cup {\cal E})$, $Im\,k=Im\,l$,
$k^2=l^2$,
$$\psi(x,k)=\psi^0(x,k)+\int_{\R^d}R^0(x,y,k)(v(y)-v^0(y))\psi(y,k)dy,
\eqno(3.2)$$
where $x\in\R^d$, $k\in\C^d\b (\R^d\cup {\cal E}^0)$ (and (3.2) at fixed $k$
is an equation for $\psi=e^{ikx}\mu(x,k)$, where $\mu$ is sought in
$L^{\infty}(\R^d)$),
$$h_{\gamma}(k,l)=h^0_{\gamma}(k,l)+\bigl({1\over 2\pi}\bigr)^d\int_{\R^d}
\psi^0_{-\gamma}(x,-l)
(v(x)-v^0(x))\psi_{\gamma}(x,k)dx \eqno(3.3)$$
for $\gamma\in\S^{d-1}$,
$k,l\in\R^d\b (0\cup {\cal E}^0_{\gamma}\cup {\cal E}_{\gamma})$,
$k^2=l^2$, $k\gamma=l\gamma$.
$$\psi_{\gamma}(x,k)=\psi^0_{\gamma}(x,k)+\int_{\R^d}R^0_{\gamma}(x,y,k)
(v(y)-v^0(y))\psi_{\gamma}(y,k)dy,\eqno(3.4)$$
where $x\in\R^d$, $\gamma\in\S^{d-1}$,
$k\in\R^d\b (0\cup {\cal E}^0_{\gamma})$ (and (3.4) at fixed $\gamma$ and
$k$ is an equation for $\psi_{\gamma}$ in $L^{\infty}(\R^d)$),
$$f(k,l)=f^0(k,l)+\bigl({1\over 2\pi}\bigr)^d\int_{\R^d}
\psi^{+,0}(x,-l)(v(x)-v^0(x))\psi^+(x,k)dx \eqno(3.5)$$
for $k,l\in\R^d\b (0\cup {\cal E}^{+,0}\cup {\cal E}^+)$, $k^2=l^2$,
$$\psi^+(x,k)=\psi^{+,0}(x,k)+\int_{\R^d}R^{+,0}(x,y,k)(v(y)-v^0(y))
\psi^+(y,k)dy, \eqno(3.6)$$
where $x\in\R^d$,
$k\in\R^d\b (0\cup {\cal E}^{+,0})$ (and (3.6) at fixed
$k$ is an equation for $\psi^+$ in $L^{\infty}(\R^d)$). (We remind that
$\psi^+$, $f$, $\psi$, $h$, $\psi_{\gamma}$, $h_{\gamma}$ were defined
in the introduction by means of (1.7)-(1.13).) Equation (3.6) is well-known
in the classical scattering theory for the Schr\"odinger equation (1.5).
Formula (3.5) was given, in particular, in [St]. To our knowledge formulas
and equations (3.1)-(3.4) were given for the first time in [No4].

Note that, under the assumption (1.6):
$$\eqalign{
&(3.2)\ \ {\rm at\ fixed}\ \ k\in\C^d\b (\R^d\cup {\cal E}^0)\ \ {\rm is\
uniquely\ solvable}\cr
&{\rm for}\ \ \psi=e^{ikx}\mu(x,k)\ \ {\rm with}\ \ \mu\in L^{\infty}(\R^d)
\ \ {\rm if\ and\ only\ if}\ \ k\notin {\cal E};\cr
&(3.4)\ \ {\rm at\ fixed}\ \ \gamma\in\S^{d-1},\ \
k\in\R^d\b (0\cup {\cal E}^0_{\gamma})\ \ {\rm is\ uniquely\ solvable}\cr
&{\rm for}\ \ \psi_{\gamma}\in L^{\infty}(\R^d)\ \ {\rm if\ and\ only\ if}\ \
 k\notin {\cal E}_{\gamma};\cr
&(3.6)\ \ {\rm at\ fixed}\ \
k\in\R^d\b (0\cup {\cal E}^{+,0})\ \ {\rm is\ uniquely\ solvable}\cr
&{\rm for}\ \ \psi^+\in L^{\infty}(\R^d)\ \ {\rm if\ and\ only\ if}\ \
k\notin {\cal E}^+.\cr}\eqno(3.7)$$

In a similar way with [No1], (under the assumptions of Theorem 1) formulas
and equations (3.1)-(3.6) can be transformed into (2.8)-(2.13) by means of
the following Green's formula
$$\eqalign{
&\int_D(u_1(x)\Delta u_2(x)-u_2(x)\Delta u_1(x))dx=\cr
&\int_{\pa D}\bigl(u_1(x){\pa u_2(x)\over \pa\nu}-u_2(x)
{\pa u_1(x)\over \pa\nu}\bigr)dx,\cr}\eqno(3.8)$$
(where $dx$ in the right-hand side of (3.8) denotes the standard measure
on $\pa D$ in $\R^d$). (To start these transformations, we use that
$\psi(x,k)$, $\psi_{\gamma}(x,k)$ and $\psi^+(x,k)$ satisfy (1.5) for
$E=k^2$ and replace $v\psi$ in (3.1), (3.2) by $(\Delta+E)\psi$,
$v\psi_{\gamma}$ in (3.3), (3.4) by $(\Delta+E)\psi_{\gamma}$ and
$v\psi^+$ in (3.5), (3.6) by $(\Delta+E)\psi^+$.)
However, these calculations can be shortened by means of the following
Alessandrini identity (being valid under the assumptions of Theorem 1):
$$\int_D(v(x)-v^0(x))\psi(x)\psi^0(x)dx=\int_{\pa D}\int_{\pa D}
\psi^0(x)(\Phi-\Phi^0)(x,y,E)\psi(y)dydx \eqno(3.9)$$
for any $\psi$ and $\psi^0$ such that $\psi$ satisfies (1.1), $\psi^0$
satisfies (1.1) with $v$ replaced by $v^0$ and where $\psi$ and $\psi^0$
are sufficiently regular in $D$, for example, $\psi,\psi^0\in H^1(D)$. In a
slightly different form the identity (3.9) was given in Lemma 1 of [A]. The
proof of (3.9) is based on (3.8).

Formula (2.8) follows from (3.1), (3.9) and the  fact that $\psi(x,k)$
satisfies (1.5) for $E=k^2$ and $\psi^0(x,-l)$ satisfies (1.5) with $v$
replaced by $v^0$, for $E=l^2$. Formulas (2.9) with $x\in\R^d\b\bar D$
follow from (3.2) with $x\in\R^d\b\bar D$, (3.9) and the fact that
$\psi(y,k)$ satisfies (1.5) in $y$ for $E=k^2$ and that $R^0(x,y,k)$ for
$x\in\R^d\b\bar D$ satisfies (1.5) in $y$ in an open neighborhood of $\bar D$,
with $v$ replaced by $v^0$, for $E=k^2$. The latter statement about $R^0$
follows from (2.7) and the symmetry
$$R(x,y,k)=R(y,x,-k),\eqno(3.10)$$
where $x,y\in\R^d$, $k\in\C^d\b\R^d$ (and $R$ is defined  by means of
(2.2)-(2.4)). The symmetry (3.10) was found in [No4]. Finally, formulas
(2.9) for $x\in\pa D$ arise as a limit of (2.9) with $x\in\R^d\b\bar D$.

The proof of (2.10)-(2.13) is similar to the proof of (2.8), (2.9).

Theorem 1 is proved.

\vskip 2 mm
{\bf Proof of Proposition 1.}
In this proof we obtain and use the following formula (being valid under
the assumption (1.6) on $v^0$ and $v$):
$$h_{\gamma}(k,l)=h^0_{\gamma}(k,l)+\bigl({1\over 2\pi}\bigr)^d
\int_{\R^d}\psi^0_{-\gamma}(x,-k,-l)(v(x)-v^0(x))
\psi_{\gamma}(x,k)dx \eqno(3.11)$$
for $\gamma\in\S^{d-1}$, $k\in\R^d\b (0\cup {\cal E}^0_{\gamma}
\cup {\cal E}_{\gamma})$, $l\in\R^d$, $k^2=l^2$. Formula (2.15) follows
from (3.11), (3.9) and the fact that $\psi_{\gamma}(x,k)$ satisfies (1.5)
for $E=k^2$ and $\psi^0_{-\gamma}(x,-k,-l)$ satisfies (1.5) with $v$
replaced by $v^0$, for $E=k^2=l^2$. Thus, to prove Proposition 1, it
remains to prove (3.11).

To prove (3.11) we use, in particular, that
$$\psi_{\gamma}(x,k,l)=\psi^0_{\gamma}(x,k,l)+\int_{\R^d}R^0_{\gamma}(x,y,k)
(v(y)-v^0(y))\psi_{\gamma}(y,k,l)dy \eqno(3.12)$$
for $x\in\R^d$, $\gamma\in\S^{d-1}$,
$k\in\R^d\b (0\cup {\cal E}^0_{\gamma}\cup {\cal E}_{\gamma})$, $l\in\R^d$,
$k^2=l^2$. To obtain (3.12) we write (2.14a) as
$$\eqalign{
&\psi_{\gamma}(x,k,l)-\int_{\R^d}G_{\gamma}(x-y,k)v^0(y)
\psi_{\gamma}(y,k,l)dy -e^{ilx}=\cr
&\int_{\R^d}G_{\gamma}(x-y,k)(v(y)-v^0(y))\psi_{\gamma}(y,k,l)dy,\cr}
\eqno(3.13)$$
where $x\in\R^d$, $\gamma\in\S^{d-1}$, $k\in\R^d\b (0\cup {\cal E}_{\gamma})$,
$l\in\R^d$, $k^2=l^2$. Replacing $e^{ilx}$ in (3.13) by its expression
from (2.14a) with $v$ and $\psi_{\gamma}$ replaced by $v^0$ and
$\psi^0_{\gamma}$ we have
$$\eqalign{
&\psi_{\gamma}(x,k,l)-\psi^0_{\gamma}(x,k,l)
-\int_{\R^d}G_{\gamma}(x-y,k)v^0(y)
(\psi_{\gamma}(y,k,l) - \psi^0_{\gamma}(y,k,l))dy=\cr
&\int_{\R^d}G_{\gamma}(x-y,k)(v(y)-v^0(y))\psi_{\gamma}(y,k,l)dy,\cr}
\eqno(3.14)$$
where $x\in\R^d$, $\gamma\in\S^{d-1}$, $k\in\R^d\b (0\cup
{\cal E}^0_{\gamma}\cup {\cal E}_{\gamma})$,
$l\in\R^d$, $k^2=l^2$.

Comparing (3.14) (as an equation for $\psi_{\gamma}-\psi^0_{\gamma}$) with
the following equation (arising from (2.2), (2.5), (2.14b)) for
$R^0_{\gamma}(x,y,k)$:
$$R^0_{\gamma}(x,y,k)-\int_{\R^d}G_{\gamma}(x-z,k)v^0(z)
R^0_{\gamma}(z,y,k)dz=G_{\gamma}(x-y,k),\eqno(3.15)$$
$x,y\in\R^d$, $\gamma\in\S^{d-1}$, $k\in\R^d\b (0\cup
{\cal E}^0_{\gamma}\cup {\cal E}_{\gamma})$, we obtain (3.12).

Note that for $v\equiv 0$ equation (3.12) with $k,\gamma$ and $l$ replaced
by $-k$, $-\gamma$ and $-l$ takes the form
$$e^{-ilx}=\psi^0_{-\gamma}(x,-k,-l)-\int_{\R^d}R^0_{-\gamma}(x,y,-k)
v^0(y)e^{-ily}dy,\eqno(3.16)$$
$x\in\R^d$, $\gamma\in\S^{d-1}$, $k\in\R^d\b (0\cup
{\cal E}^0_{\gamma})$. To prove (3.11) we will use also the following
symmetry (following from (3.10))
$$R_{\gamma}(x,y,k)=R_{-\gamma}(y,x,-k),\ \ x,y\in\R^d,\ \ \gamma\in\S^{d-1},
\ \ k\in\R^d\b 0.\eqno(3.17)$$

The following sequences of equalities proves (3.11):
$$\eqalign{
&(2\pi)^dh_{\gamma}(k,l)=
\int_{\R^d}e^{-ilx}v(x)\psi_{\gamma}(x,k)dx=
\int_{\R^d}e^{-ilx}v^0(x)\psi_{\gamma}(x,k)dx+\cr
&\int_{\R^d}e^{-ilx}(v(x)-v^0(x))\psi_{\gamma}(x,k)dx\buildrel (3.16) \over =
\int_{\R^d}e^{-ilx}v^0(x)\psi_{\gamma}(x,k)dx+\cr
&\int_{\R^d}\psi^0_{-\gamma}(x,-k,-l)(v(x)-v^0(x))\psi_{\gamma}(x,k)dx-\cr
&\int_{\R^d}\int_{\R^d}R^0_{-\gamma}(x,y,-k)v^0(y)e^{-ily}
(v(x)-v^0(x))\psi_{\gamma}(x,k)dydx\buildrel (3.17) \over =\cr
&\int_{\R^d}e^{-ilx}v^0(x)\psi_{\gamma}(x,k)dx+
\int_{\R^d}\psi^0_{-\gamma}(x,-k,-l)(v(x)-v^0(x))\psi_{\gamma}(x,k)dx-\cr
&\int_{\R^d}\int_{\R^d}R^0_{\gamma}(y,x,k)v^0(y)e^{-ily}
(v(x)-v^0(x))\psi_{\gamma}(x,k)dxdy\buildrel (3.4) \over =\cr
&\int_{\R^d}e^{-ilx}v^0(x)\psi_{\gamma}(x,k)dx+
\int_{\R^d}\psi^0_{-\gamma}(x,-k,-l)(v(x)-v^0(x))\psi_{\gamma}(x,k)dx+\cr
&\int_{\R^d}e^{-ily}v^0(y)(\psi^0_{\gamma}(y,k)-\psi_{\gamma}(y,k))dy=\cr
&(2\pi)^dh^0_{\gamma}(k,l)+
\int_{\R^d}\psi^0_{-\gamma}(x,-k,-l)(v(x)-v^0(x))\psi_{\gamma}(x,k)dx\cr}$$
for $\gamma\in\S^{d-1}$, $k\in\R^d\b (0\cup {\cal E}^0_{\gamma}\cup
{\cal E}_{\gamma})$, $l\in\R^d$, $k^2=l^2$.

Proposition 1 is proved.

\vskip 2 mm
{\bf Proof of Proposition 2.}
Let us prove Proposition 2 for the case of equation (2.9a). Consider the
operator $A(k)$ of (2.9a):
$$A(k)=R^0(k)\,(\Phi(E)-\Phi^0(E)),\eqno(3.18)$$
where $k\in\C^d\b (\R^d\cup {\cal E}^0)$, $k^2=E$, and where the operator
$R^0(k)$ is defined by
$$R^0(k)\v(x)=\int_{\pa D}R^0(x,y,k)\v(y)dy,\ \ x\in\pa D,\eqno(3.19)$$
where $\v$ is a test function.

Under the assumptions of Theorem 1, the operator $\Phi(E)-\Phi^0(E)$ is
compact in $L^{\infty}(\pa D)$. This follows from the following properties
of the Schwartz kernel of  $\Phi(E)-\Phi^0(E)$:
$$(\Phi-\Phi^0)(x,y,E)\ \ {\rm is\ continuous\ for}\ \ x,y\in\pa D,\ \
x\ne y,\eqno(3.20)$$
$$\eqalign{
&|(\Phi-\Phi^0)(x,y,E)|\le C_1|x-y|^{2-d},\ \ x,y\in\pa D,\ \ {\rm for}\ \
d\ge 3,\cr
&|(\Phi-\Phi^0)(x,y,E)|\le C_1|\ln\,|x-y||,\ \ x,y\in\pa D,\ \ {\rm for}\ \
d=2,\cr}\eqno(3.21)$$
where $C_1$ is some constant (dependent on $D$, $v$, $v^0$, $E$ and $d$).
Note that for $v^0\equiv 0$ the result that $\Phi(E)-\Phi^0(E)$ is compact
in $L^{\infty}(\pa D)$ (under the assumptions of Theorem 1) was given in
[No1].

If (1.2a) is fulfilled, $v^0$ satisfies (1.15) and
$k\in\C^d\b (\R^d\cup {\cal E}^0)$, then $R^0(k)$ is a compact operator
in $L^{\infty}(\pa D)$. This follows already from the following
properties of $R^0(x,y,k)$:
$$R^0(x,y,k)\ \ {\rm is\ continuous\ for}\ \ x,y\in\bar D,\ \ x\ne y,
\eqno(3.22)$$
$$\eqalign{
&|R^0(x,y,k)|\le C_2|x-y|^{2-d},\ \ x,y\in\bar D,\ \ {\rm for}\ \ d\ge 3,\cr
&|R^0(x,y,k)|\le C_2|\ln\,|x-y||,\ \ x,y\in\bar D,\ \ {\rm for}\ \ d=2,\cr}
\eqno(3.23)$$
where $C_2$ is some constant (dependent on $D$, $v^0$, $k$ and $d$).

Actually, under the aforementioned assumptions on $D$, $v^0$ and $k$,
$R^0(k)$ is a bounded operator from $L^{\infty}(\pa D)$ to
$C^{\alpha}(\pa D)$ for any $\alpha\in [0,1[$ (where $C^{\alpha}$ denotes
the H\"older space). This result in the general case follows from this
result for $v^0\equiv 0$ (when $R^0(x,y,k)=G(x-y,k)$), the relation (2.2),
 the estimate (3.23) and the property that
$$\eqalign{
&\int_DG(x-z,k)u(z)dz\in C^{\alpha}(\R^d),\ \ {\rm at\ least,\ for\ any}\ \
\alpha\in [0,1]\cr
&{\rm (as\ a\ function\ of}\ x)\ \ {\rm for}\ \ u\in L^{\infty}(D)\ \
{\rm (and}\ \ k\in\C^d\b\R^d).\cr}$$

The formula (3.18) and the aforementioned properties of the  operators
 $\Phi(E)-\Phi^0(E)$ and $R^0(k)$ imply that, under the assumptions
of Theorem 1, for fixed $k\in\C^d\b (\R^d\cup {\cal E}^0)$, $k^2=E$,
$A(k)$ is a compact operator in $L^{\infty}(\pa D)$ and, thus, (2.9a) is a
Fredholm linear integral equation of the second kind for $\psi$ in
$L^{\infty}(\pa D)$.

Under the assumptions of Theorem 1, for $k\in\C^d\b (\R^d\cup {\cal E}^0)$,
$k^2=E$, the aforementioned properties of $\Phi(E)-\Phi^0(E)$ and
$R^0(k)$ and the property that $\psi^0\in C^{\alpha}(\pa D)$,
$\alpha\in [0,1[$, imply also that
$$
{\rm if}\ \ \psi\in L^{\infty}(\pa D)\ \ {\rm satisfies}\ \ (2.9a),\ \
{\rm then}\ \ \psi\in C^{\alpha}(\pa D)\ \ {\rm for\ any}\ \
\alpha\in [0,1[.\eqno(3.24)$$

Note also that if (1.2a) is fulfilled, $v^0$ satisfies (1.15) and
$k\in\C^d\b (\R^d\cup {\cal E}^0)$, then $R^0(k)$ is a bounded
operator from $L^2(\pa D)$ to $H^1(\pa D)$. One can prove this result in
the general case proceeding from this result for $v^0\equiv 0$ (when
$R^0(x,y,k)=G(x-y,k)$), equation (2.2) with its iterations for
$R(x,y,k)$, $R(x,y,k)-G(x-y,k)$ and so on, estimate (3.23) and the property
that
$$\int_DG(x-z,k)u(z)dz\in H^2(D)\ \ {\rm (as\ a\ function\ of}\ \ x)\ \
{\rm for}\ \ u\in L^2(D)\ \ {\rm (and}\ \ k\in\C^d\b\R^d).$$
(For $v^0\equiv 0$ this result was given, for example, in [Na1].)
Therefore, under the assumptions of Theorem 1, for
$k\in\C^d\b (\R^d\cup {\cal E}^0)$, $k^2=E$, one can see that
$${\rm if}\ \ \psi\in L^{\infty}(\pa D)\ \ {\rm satisfies}\ \ (2.9a),\ \
{\rm then\ also}\ \ \psi\in H^1(\pa D).\eqno(3.25)$$

To prove Proposition 2 for the case of equation (2.9a) it remains to show
that, under the assumptions of Theorem 1, equation (2.9a) at fixed
$k\in\C^d\b (\R^d\cup {\cal E}^0)$, $k^2=E$, is uniquely solvable for
$\psi\in L^{\infty}(\pa D)$ iff $k\not\in {\cal E}$.

For  $k\in\C^d\b (\R^d\cup {\cal E}^0)$, under the assumption (1.6) for
$v$ and $v^0$, due to (2.1a) and (3.7), $k\not\in {\cal E}$ iff
equation (3.2) is uniquely solvable for $\psi=e^{ikx}\mu(x,k)$ with
$\mu\in L^{\infty}(\R^d)$. In turn, for
$k\in\C^d\b (\R^d\cup {\cal E}^0)$, under the assumption (1.15) for
$v$ and $v^0$,
equation (3.2) is uniquely solvable for $\psi=e^{ikx}\mu(x,k)$ with
$\mu\in L^{\infty}(\R^d)$ iff
 (3.2) is uniquely solvable for $\psi\in C(\R^d)$.
Thus, it remains to show that, under the assumptions of Theorem 1, equation
(2.9a) at fixed
$k\in\C^d\b (\R^d\cup {\cal E}^0)$, $k^2=E$, is uniquely solvable for
$\psi\in L^{\infty}(\pa D)$ iff (3.2) is uniquely solvable for
$\psi\in C(\R^d)$. This proof consists of the following two parts.

\vskip 2 mm
Part 1.
Suppose that (under the assumptions of Theorem 1) at fixed
$k\in\C^d\b (\R^d\cup {\cal E}^0)$, $k^2=E$, equation (3.2) has several
solutions in $C(\R^d)$. Then, repeating the proof of (2.9) separately
for each solution, we find that the restriction to $\pa D$ of each of
these solutions satisfies (2.9a). In addtion, different solutions
$\psi$ have different restrictions to $\pa D$. This follows from (1.3).
Thus at fixed
$k\in\C^d\b (\R^d\cup {\cal E}^0)$, $k^2=E$, equation (2.9a) has, at least,
as many solutions as equation (3.2).

\vskip 2 mm
Part 2.
To prove the converse, we use the following identities:
$$\eqalign{
&\int_DR^0(x,y,k)(v(y)-v^0(y))\psi(y)dy\buildrel (1.1) \over =
\int_DR^0(x,y,k)(\Delta+E)\psi(y)dy-\cr
&\int_DR^0(x,y,k)v^0(y)\psi(y)dy\buildrel (3.8) \over =
\int_D\psi(y)(\Delta_y+E-v^0(y))R^0(x,y,k)dy+\cr
&\int_{\pa D}\bigl(R^0(x,y,k){\pa\over \pa \nu_y}\psi(y)-
\psi(y){\pa\over \pa \nu_y}R^0(x,y,k)\bigr)dy\buildrel (2.7),(3.10) \over =
\cr
&\int_D\psi(y)\delta(x-y)dy+
\int_{\pa D}\bigl(R^0(x,y,k){\pa\over \pa \nu_y}\psi(y)-
\psi(y){\pa\over \pa \nu_y}R^0(x,y,k)\bigr)dy\cr
&{\rm for}\ \ x\in\R^d\b\pa D,\cr}\eqno(3.26)$$

$$\int_DR^0(x,y,k)(v(y)-v^0(y))\psi(y)dy\buildrel (3.9),(2.9b) \over =
\int_{\pa D}A(x,y,k)\psi(y)dy\ \ {\rm for}\ \ x\in\R^d\b\bar D,\eqno(3.27)$$
where  $k\in\C^d\b (\R^d\cup {\cal E}^0)$, $k^2=E$, $\psi$ satisfies (1.1)
and is sufficiently regular in $D$, for example, $\psi\in H^1(D)$.

Let  $k\in\C^d\b (\R^d\cup {\cal E}^0)$, $k^2=E$, be fixed. Suppose that
$\psi\in L^{\infty}(\pa D)$ solves (2.9a). Due to (3.24), (3.25), we
have also that $\psi\in C^{\alpha}(\pa D)$, $\alpha\in [0,1[$, and
$\psi\in H^1(\pa D)$. Consider this $\psi$ as Dirichlet data for
equation (1.1) and consider the solution $\psi$ (of (1.1)) corresponding
to these data. We have that $\psi\in C^{\alpha}(\bar D)$, $\alpha\in
[0,1[$, and $\psi\in H^{3/2}(D)$.

Let $\psi$ be also defined on $\R^d\b\bar D$ by (2.9) with
$x\in\R^d\b\bar D$ (in terms of $\psi\big|_{\pa D}$). Let us prove that $\psi$
defined in such a way on $\R^d=\pa D\cup D\cup (\R^d\b\bar D)$
satisfies (3.2) and belongs to $C(\R^d)$. As a particular case of the
aforementioned property $\psi\in C^{\alpha}(\bar D)$, $\alpha\in [0,1[$,
we have that $\psi\in C(\bar D)$. Proceeding from the definition of $\psi$
on $\R^d\b\bar D$, one can easily show that, at least,
$\psi\in C(\R^d\b D)$. The properties $\psi\in C(\bar D)$ and
$\psi\in C(\R^d\b D)$ imply that $\psi\in C(\R^d)$. The proof that $\psi$
satisfies (3.2) consists in the following. First, from (2.9) with
$x\in\R^d\b\bar D$ and (3.27) (and the continuity of $\psi$) we obtain
that $\psi$ satisfies (3.2) for $x\in\R^d\b D$. In addition, taking into
account (3.26) we have also that
$$\psi(x)=\psi^0(x,k)+\int_{\pa D}\bigl(R^0(x,y,k){\pa\over \pa \nu_y}\psi(y)
- \psi(y){\pa\over \pa \nu_y}R^0(x,y,k)\bigr)dy\eqno(3.28)$$
for $x\in\R^d\b\bar D$, where ${\pa\over \pa \nu}\psi$ is taken for $\psi$
defined on $\bar D$. Further, (as well as for $v^0\equiv 0$, see
[No1],[Na2]) proceeding from (3.28) we obtain that
$$\psi(x)=\psi^0(x,k)+\psi(x)+
\int_{\pa D}\bigl(R^0(x,y,k){\pa\over \pa \nu_y}\psi(y)
- \psi(y){\pa\over \pa \nu_y}R^0(x,y,k)\bigr)dy\eqno(3.29)$$
for $x\in D$, where ${\pa\over \pa \nu}\psi$ is taken for $\psi$ defined
in $\bar D$. Note that proceeding from (3.28) and using that
$\psi\in H^1(\pa D)$, ${\pa\over \pa \nu}\psi\in L^2(\pa D)$ and
the jump property of the double lager potential ${\pa\over \pa \nu_y}
R^0(x,y,k)$, we obtain, first, (3.29) in the limit for $x=\xi-0\nu_{\xi}$,
$\xi\in\pa D$ (where ${\nu}_{\xi}$ is the outward normal to $\pa D$ at
$\xi$). Further, using that $\psi^0(x,k)$ and $R^0(x,y,k)$ with
${\pa\over \pa\nu_y}R^0(x,y,k)$, $y\in\pa D$, satisfy (1.1) with $v^0$ in
place of $v$, for $E=k^2$, we obtain (3.29) for $x\in D$. Finally,
from (3.29) and (3.26) we obtain that $\psi$ satisfies (3.2) also for
$x\in D$. Thus, any solution $\psi$ of (2.9a) can be continued to a
continuous solution of (3.2). This completes the (part 2 of) proof that,
under the assumptions of Theorem 1, at fixed
$k\in\C^d\b(\R^d\cup {\cal E}^0)$, $k^2=E$, equation (2.9a) is uniquely
solvable for $\psi\in L^{\infty}(\pa D)$ iff (3.2) is uniquely
solvable for $\psi\in C(\R^d)$.

The proof of Proposition 2 for the case of equation (2.9a) is completed.
The proof of Proposition 2 for the cases of equations (2.11a) and
(2.13a) is similar.

\vskip 10 mm
{\bf References}
\vskip 4 mm
\item{[    A]} Alessandrini G 1988 Stable determination of conductivity
by boundary measurements {\it Appl. Anal.} {\bf 27} 153-172
\item{[   BC]} Beals R and Coifman R R 1985 Multidimensional inverse
scattering and nonlinear partial differential equations {\it Proc. Symp.
Pure Math.} {\bf 43} 45-70
\item{[BBMRS]} Bogatyrev A V, Burov V A, Morozov S A, Rumyantseva O D
and Sukhov E G  2000 Numerical realization of algorithm for exact solution of
two-dimensional monochromatic inverse problem of acoustical scattering
{\it Acoustical Imaging} {\bf 25}  (Kluwer Academic/Plenum Publishers,New York)
\ 65-70
\item{[  BMR]} Burov V A, Morozov S A and Rumyantseva O D 2002
Reconstruction of fine-scale structure of acoustical scatterer on large-scale
contrast background
{\it Acoustical Imaging} {\bf 26} (Kluwer Academic/Plenum Publishers,New York)
\ 231-238
\item{[   Ch]} Chen Yu 1997 Inverse scattering via Heisenberg's
uncertainty principle {\it Inverse Problems} {\bf 13} 253-282
\item{[    E]} Eskin G 2001 The inverse scattering problem in two dimensions
at a fixed energy {\it Comm. Partial Differential Equations}  {\bf 26}
1055-1090
\item{[   F1]} Faddeev L D 1965 Increasing solutions of the Shr\"odinger
equation {\it Dokl. Akad. Nauk SSSR} {\bf 165} 514-517 (in Russian);
English transl.: 1966 {\it Sov. Phys. Dokl.} {\bf 10} 1033-1035
\item{[   F2]} Faddeev L D 1966 Factorization of the $S$ matrix for the
multidimensional Shr\"odinger operator {\it Dokl. Akad. Nauk SSSR}
{\bf 167} 69-72 (in Russian); English transl.: 1966 {\it Sov. Phys. Dokl.}
{\bf 11} 209-211
\item{[   F3]} Faddeev L D 1974 Inverse problem of quantum scattering theory
II {\it Itogi Nauki Tekhn. Sovrem. Prob. Mat.} {\bf 3} 93-180 (in Russian);
English transl.: 1976 {\it Sov. Math.} {\bf 5} 334-396
\item{[   GN]} Grinevich P G and Novikov R G 1995 Transparent potentials
at fixed energy in dimension two. Fixed-energy dispersion relations for the
fast decaying potentials {\it Comm. Math. Phys.} {\bf 174} 409-446
\item{[   HN]} Henkin G M and Novikov R G 1987 The $\bar\pa$- equation in
the multidimensional inverse scattering problem {\it Uspekhi Mat. Nauk}
{\bf 42} 93-152 (in Russian); English transl.: 1987 {\it Russ. Math. Surv.}
{\bf 42} 109-180
\item{[   IS]} Isakov V and Sun Z 1995 The inverse scattering at fixed
energies in two dimensions {\it Indiana Univ. Math. J.} {\bf 44} 883-896
\item{[   LN]} Lavine R B and Nachman A I 1987 On the inverse scattering
transform of the $n$- dimensional Schr\"odinger operator In {\it Topics in
Soliton Theory and Exactly Solvable Nonlinear Equations (M.Ablovitz,
B.Fuchssteiner and M.Kruskal, Eds.), pp. 33-44. World Scientific, Singapore}
\item{[    M]} Mandache N 2001 Exponential instability in an inverse
problem for the Schr\"odinger equation {\it Inverse Problems} {\bf 17}
1435-44
\item{[  Na1]} Nachman A I 1988 Reconstructions from boundary
measurements {\it Ann. Math.} {\bf 128} 531-576
\item{[  Na2]} Nachman A I 1995 Global uniqueness for a two-dimensional
inverse boundary value problem {\it Ann. Math.} {\bf 142} 71-96
\item{[  NSU]} Nachman A I, Sylvester J and Uhlmann G 1988 An $n$-
dimensional Borg-Levinson theorem {\it Comm. Math. Phys.} {\bf 115}
593-605
\item{[   Ne]} Newton R G 1989 Inverse Schr\"odinger scattering in three
dimensions {\it Springer-Verlag,}

\noindent
\item{       } {\it Berlin}
\item{[  No1]} Novikov R G 1988 Multidimensional inverse spectral problem
for the equation $-\Delta\psi+(v(x)-Eu(x))\psi=0$. {\it Funkt. Anal. Ego
Pril} {\bf 22} 11-22 (in Russian); English transl.: 1988 {\it Funct. Anal.
Appl.} {\bf 22} 263-272
\item{[  No2]} Novikov R G 1992 The inverse scattering problem on a fixed
energy level for the two-dimensional Schr\"odinger operator {\it J. Funct.
Anal.} {\bf 103} 409-463
\item{[  No3]} Novikov R G 1994 The inverse scattering problem at fixed
energy for the  three-dimensional Schr\"odinger equation with an
exponentially decreasing potential {\it Comm. Math. Phys.} {\bf 161}
569-595
\item{[  No4]} Novikov R G 1996 $\bar\pa$-method with nonzero background
potential. Application to inverse scattering for the two-dimensional
acoustic equation {\it Comm. Partial Differential Equations} {\bf 21}
597-618
\item{[  No5]} Novikov R G 1999 Approximate inverse quantum scattering
at fixed energy in dimension 2 {\it Proc. Steklov Inst. of Math.}
{\bf 225} 285-302
\item{[  No6]} Novikov R G 2002 Scattering for the Schr\"odinger equation
in multidimension. Non-linear $\bar\pa$-equation, characterization of
scattering data and related results. {\it Chapter 6.2.4 in SCATTERING
edited by E.R.Pike and P.Sabatier,  Academic Press}
\item{[   St]} Stefanov P 1990 A uniqueness result for the inverse
back-scattering problem {\it Inverse Problems} {\bf 6} 1055-1064
\item{[   SU]} Sylvester J and Uhlmann G. 1987 A global uniqueness theorem
for an inverse boundary value problem {\it Ann. of Math.} {\bf 125}
153-169
\item{[   We]} Weder R 1991 Generalized limiting absorption method and
multidimensional inverse scattering theory {\it Math. Methods Appl. Sci.}
{\bf 14} 509-524

\end